\documentclass[sigconf]{acmart}

\usepackage{booktabs,multirow} 
\usepackage{hyperref}
\usepackage{amsmath}
\usepackage[ruled,vlined,linesnumbered]{algorithm2e}
\usepackage{soul}
\usepackage{bm}
\usepackage{enumitem}
\usepackage{tikz}

\makeatletter 
\def\algbackskip{\hskip-\ALG@thistlm}
\makeatother 

\begin{document}

\title[Results of QTA on the ATSP]{Focusing on the Hybrid Quantum Computing - Tabu Search Algorithm: new results on the Asymmetric Salesman Problem}


\author{Eneko Osaba}
\orcid{0000-0001-7863-9910}
\affiliation{%
	\institution{TECNALIA, Basque Research and \\ Technology Alliance (BRTA)\\20009, Donostia-San Sebastian, Spain}
}
\email{eneko.osaba@tecnalia.com}

\author{Esther Villar-Rodriguez}
\orcid{}
\affiliation{%
     \institution{TECNALIA, Basque Research and \\ Technology Alliance (BRTA)\\20009, Donostia-San Sebastian, Spain}
}
\email{esther.villar@tecnalia.com}

\author{Izaskun Oregi}
\orcid{}
\affiliation{%
	\institution{TECNALIA, Basque Research and \\ Technology Alliance (BRTA)\\20009, Donostia-San Sebastian, Spain}
}
\email{izaskun.oregui@tecnalia.com}

\author{Aitor Moreno-Fernandez-de-Leceta}
\orcid{}
\affiliation{%
	\institution{Instituto Ibermatica de Innovacion. \\ Parque Tecnologico de Alava, \\01510 Mi\~nano, Spain}
}
\email{ai.moreno@ibermatica.com}

\renewcommand{\shortauthors}{E. Osaba et al.}



\begin{abstract}
Quantum Computing is an emerging paradigm which is gathering a lot of popularity in the current scientific and technological community. Widely conceived as the next frontier of computation, Quantum Computing is still at the dawn of its development. Thus, current solving systems suffer from significant limitations in terms of performance and capabilities. Some interesting approaches have been devised by researchers and practitioners in order to overcome these barriers, being quantum-classical hybrid algorithms one of the most often used solving schemes. The main goal of this paper is to extend the results and findings of the recently proposed hybrid \textit{Quantum Computing - Tabu Search} Algorithm for partitioning problems. To do that, we focus our research on the adaptation of this method to the Asymmetric Traveling Salesman Problem. In overall, we have employed six well-known instances belonging to TSPLIB to assess the performance of \textit{Quantum Computing - Tabu Search} Algorithm in comparison to QBSolv, a state-of-the-art decomposing solver. Furthermore, as an additional contribution, this work also supposes the first solving of the Asymmetric Traveling Salesman Problem using a Quantum Computing based method. Aiming to boost whole community's research in QC, we have released the project's repository as open source code for further application and improvements. 
\end{abstract}

\keywords{Quantum Computing, Metaheuristic Optimization, Traveling Salesman Problem, Transfer Optimization, DWAVE}

\maketitle


\section{Introduction} \label{sec:intro}

Quantum Computing (QC, \cite{steane1998quantum}) is deemed as the next frontier in computation. The current scientific community has high expectations on this specific paradigm, being this the main reason for its fast-growing popularity. Among all its potential benefits, QC has arisen as a promising alternative for solving optimization problems. Thus, QC provides a revolutionary approach for tackling this kind of tasks, offering clear theoretical advantages ranging from significant computation speed to efficient search ability \cite{ajagekar2020quantum}. 

Two QC architecture types are available today: annealing-based quantum computers and gate-model quantum devices \cite{wang201816,hauke2020perspectives}. This paper is devoted to the first of these architectures, which is most used one for optimization purposes.

Although scientific community has devoted many efforts in recent years to the advancement of QC paradigm, QC is still at the dawn of its development. Moreover, current commercial QC systems suffer from huge computational and performance limitations \cite{fellous2020limitations,al2017natural}. In this paper, we focus our attention on the quantum-annealer provided by D-Wave Systems \cite{yin2017quantum}, which has emerged as the most employed commercial QC device on these days. In any case, even this architecture has its own drawbacks despite being the most used QC machine in the current community \cite{gibney2017d}. Problems such as decoherence, limited control to quantum resources or poor error correction suppose unavoidable obstacles for researchers and practitioners for the formulation of efficient and reliable purely quantum solving approaches. 

The objective of overcoming current limitations of QC have led researchers to the design and development of hybrid quantum-classical hybrid solving methods. A representative example of these approaches can be found in \cite{feld2019hybrid}.This paper proposes a hybrid technique for solving the Capacitated Vehicle Routing Problem using D-Wave Quantum Annealer. Another quantum-classical scheme is described in \cite{clark2019towards} for solving multi-robot routing on a grid in real time. Specifically, the implementation counts on a classical-based mechanism for generating candidate paths and a D-Wave quantum-annealing-based strategy for choosing the optimal combination of paths. In the recent \cite{ajagekar2020quantum} a hybrid technique is proposed for dealing with large-scale optimization problems, both continuous and discrete. The major asset of that paper is its extensive experimentation, focused on four heterogeneous applications: vehicle routing, manufacturing cell formation, job-shop scheduling and molecular formation. Additional remarkable work is also present in \cite{sweke2020stochastic} with the development of a quantum-classical gradient descent-based optimizer.

In this context, authors of this paper have recently proposed in \cite{osaba2020hybrid} a Hybrid Quantum Computing - Tabu Search Algorithm (QTA) for solving Partitioning Problems. Thus, the main objective of this paper is to solidify the preliminary concepts described in that paper. To do that, first we describe our previously introduced method. Then, we investigate its application on the well-known Asymmetric Traveling Salesman Problem (ATSP, \cite{oncan2009comparative,cirasella2001asymmetric}). In a nutshell, the ATSP is characterized by the particularity of having asymmetric costs, which means that traveling from a node $i$ to another node $j$ may not equal the reverse trip. Despite being a feature that bringing realism and complexity to the optimization problem, asymmetric costs have been historically less studied by the community. With all this, the main contributions of this work are threefold:

\begin{itemize}
	\item In this work, we elaborate on the application of our QTA on another partitioning problem: the ATSP. Also interesting is the size of the instances considered in this work. Up to now, most investigations focused on solving discrete optimization problems through QC deal with very small instances. In this paper we dedicate our efforts to the facing of bigger instances. More concretely, we conduct an experimentation using six different instances obtained from the well-known TSPLIB Benchmark \cite{reinelt1991tsplib}, composed by 17 to 48 nodes. Although this size is indeed small when comparing with classical computing ATSP solvers, they are much larger than the instances often included in QC research.

	\item As far as we are concerned, this work supposes the first solver of the ATSP using a QC-based method. Through this contribution, we attend some recent calls made by the community in papers such as \cite{warren2020solving} and \cite{warren2017small}. These studies explicitly evince the scientific interest in ATSP instances extracted from the TSPLib. In line with this, and with the aim of strengthening the value of this work, we release the whole developed source code which embraces mechanisms for dealing with both TSP and ATSP instances. Our final goal with this initiative is to favor the advancement of the related field.
	
	\item Lastly, by using our QTA we also contribute to the field of Transfer Optimization (TO, \cite{gupta2017insights}). Specifically, we employ a multiform multitasking technique as initialization procedure for the QTA. The algorithm used for this purpose is a multiform variant of the Coevolutionary Variable Neighborhood Search Algorithm (CoVNS) for Discrete Multitasking proposed in \cite{osaba2020CoVNS}.

\end{itemize}

The remainder of this work is organized as follows: we introduce some background related to both QC and ATSP in Section \ref{sec:back}. In Section \ref{sec:QTA}, we briefly detail the main features of our QTA. After that, we describe the experimentation carried out in Section \ref{sec:exp} along with the results and discussion. We finish this paper in Section \ref{sec:conc} with main conclusions and further work. 

\section{Background} \label{sec:back}

As stated in the introduction, this section is meant to provide a brief background on two main concepts studied in this paper: Quantum Computing (Section \ref{sec:quantum}) and the ATSP (Section \ref{sec:ATSP}).

\subsection{Quantum Computing} \label{sec:quantum}

QC is gaining an increasing popularity, even breaking the barriers of scientific-world and bringing the attention of many generalist newscasts and magazines. As aforementioned, the optimization of complex problems is one of the fields expected to benefit from the advances of QC \cite{nielsen2002quantum,ayub2020quantum}.

Quantum computers are specific machines which can perform computation by means of quantum mechanical phenomena as entanglement and superposition. These machines work with a unit coined as quantum bit, or \textit{qubit} \cite{hey1999quantum}. In a nutshell, a qubit is conceived as the quantum version of the classical bit, being able to store much more information by virtue of its superposition condition. Precisely, a qubit can simultaneously depict properties of both 0 and 1 \cite{clark2019towards}, surmounting the obstacle of classical binary representations. More concretely, the superposition of qubits favors the possibility of deeming infinite quantum states before collapsing into a classical basic state (0 or 1) \cite{ajagekar2020quantum}. A further essential feature for properly understanding QC potential is the so-called entanglement. Through this characteristic, states may encode correlations between different particles, allowing for interactions and subsequent influences among chained qubits.

As commented, two kinds of QC architectures can be currently defined: gate-model quantum computers and the annealing-based quantum computers. On the one hand, gate-model quantum computers are featured by employing quantum gates for the state manipulation of the qubits. The operation of these gates can be slightly compared with the traditional logic gates, and they are applied to qubit states in a sequential way, evolving them up to their final state \cite{gyongyosi2018quantum}. Today, commercial devices have from 10 to 50 qubits, and some significant applications are search methods (Grover's Algorithm) \cite{grover1997quantum}, optimization problems \cite{farhi2014quantum} or integer factorization (Shor's Algorithm) \cite{shor1999polynomial}.

On other hand, the main rationale behind annealing-based quantum computers is the search of a minimum-energy state of a given quantum Hamiltonian (i.e., energy function). To do that, this sort of devices employ quantum related concepts such as entanglement and superposition. On this regard, the Hamiltonian is implemented as the objective function of the optimizing problem, representing its optimal solution the lowest energy state  \cite{lucas2014ising}. Today, the leading provider company for this type of device is the Canadian D-WAVE, which currently counts with the named as \textit{D-Wave Advantage\_system1.1} computer. This machine, located in Vancouver and openly accessible though LEAP cloud service\footnote{https://www.dwavesys.com/take-leap}, works with a graph composed by 5436 qubits distributed in a Pegasus topology. Specifically, this topology improves the previously one based on the well-known Chimera. Thus, each qubit is connected to other 15 qubits, instead of being linked to 6, as in previous versions of D-Wave quantum devices. Thanks to this recently deployed feature, \textit{D-Wave Advantage\_system1.1} is the most connected of any commercial quantum device in the world\footnote{https://www.dwavesys.com/press-releases/d-wave-previews-next-generation-quantum-computing-platform}.

Going deeper, in this QC machine the Hamiltonian is represented as an Ising model $\mathcal{H}_{\text{Ising}} = \sum_{i=1}^n h_i q_i + \sum_{j>i} J_{i,j} q_i q_j,$ where $q_i\in\{-1, 1\}$ depicts the $i$-th qubit, $h_i$ represents the linear bias associated to this variable, and $J_{i,j}$ is the coupling strength between $q_i$ and $q_j$ qubits. Furthermore, D-Wave QC devices allows its users to use an alternative mathematical formulation, based on the well-known Quadratic Unconstrained Binary Optimization (QUBO, \cite{glover2018tutorial}). QUBO can be formally expressed as:
\begin{equation}
\mathbf{z}^* = \min_{\mathbf{z}\in\{0,1\}^{n}} \mathbf{z}^T \cdot \mathbf{Q} \cdot \mathbf{z}.
\end{equation} Where $\mathbf{Q}$ is a programmable upper triangular matrix containing the couplers and the bias needed by the Ising model. Specifically, the nonzero off-diagonal elements represent the coupling coefficients while diagonal terms are the linear biases. In this way, $\mathbf{z}$ depicts a $n$-length binary variable array, and $\mathbf{z}^*$ the state that minimizes the quadratic function, namely, the problem solution.

As pointed before, limitations of the current QC devices, have led the community to rely on hybrid models as good alternatives for solving complex problems \cite{franca2020limitations}. 

We have spotlighted some of these alternatives in the introduction of this paper, but many more can be found in the literature. The study proposed in \cite{chiscop2020hybrid}, for example, is devoted to the Multi-Service Location Set Covering Problem using a hybrid technique considering D-Wave as the QC component. Authors in \cite{li2017hybrid} propose a quantum-classical hybrid solver for facing the quantum optimal control problem whereas a research on community detection on graphs is conducted in \cite{negre2020detecting}. Hybrid classical-quantum methods can also be found in \cite{fernandez2020hybrid} and \cite{bravyi2020hybrid} for operating on financial index tracking and graph coloring problems, respectively. Further examples of hybrid solvers can be found in \cite{guerreschi2017practical,gentini2019noise,nannicini2019performance}. Finally, we want to highlight the work proposed by Warren in 2020 in \cite{warren2020solving}, which is strictly related to the research we are describing in this paper. In that study, Warren accurately outlines the current state of the QC research community regarding the TSP. Also, the author poses some challenges and opportunities which should guide future efforts to be made by the community. Among these challenges, the solving of ATSP on a QC device is incorporated.

It is also appropriate to mention in this section the solving tool known as QBSolv. This strategy is the most employed one by the scientific community for solving large-sized optimization problems. It consists on an algorithm able to overcome the current D-Wave hardware limitations. Thus, D-Wave can face optimization problems with a higher size. Some interesting examples of the application of this tool can be seen in \cite{saito2020quantum,hussain2020optimal,mniszewski2020downfolding,okada2019improving}. In brief, QBSolv is a partitioning solving tool offered by D-Wave for splitting large complete QUBO into smaller-sized subQUBOs, tackling them in an independent and sequential way. More specifically, QBSolve breaks the complete QUBO matrix using a tabu search heuristic, and it can be run in a pure local way or using the D-Wave hardware. In any case, and despite being a promising method, QBSolv has its own limitations, as highlighted in recent studies such as \cite{teplukhin2020electronic}. One of these limitations is the high usage of the QC resources. Besides, and despite that it is slightly configurable, the fine tuning of QBSolv demands a significant access to quantum services. This is so because it is not possible for the user to fully control the usage of the D-Wave QC hardware. We refer interested readers on this tool to \cite{booth2017partitioning}.

\subsection{Asymmetric Traveling Salesman Problem as Benchmarking Problem}\label{sec:ATSP}

As introduced before, this work supposes the first solving of the well-known ATSP employing QC technologies. Being a specific variant of the canonical Traveling Salesman Problem (TSP), the ATSP can be formulated as a complete graph $G=(V,A)$, where $V= \{v_1,v_2,\dots,v_N\}$ represents the group of $N=|V|$ nodes, and $A= \{(v_i,v_j): (v_i,v_j)  \in V^2 \text{ for }i\neq j \}$ the set of arcs interconnecting these vertices. Moreover, each connection has an associated $c_{ij}$ cost, which in this case differs from its reverse path, that is $c_{ij}$ $\neq$ $c_{ji}$. In this way, the main goal of the ATSP is to calculate a complete route $TSP^*$ visiting each node only once, minimizing the total cost and finishing at the same point in which the route starts.

Despite having some crucial advantages in comparison to the basic TSP, such as a more realistic nature, ATSP has been less studied by the community along the years. Nonetheless, many remarkable articles can be found in the literature using this specific problem. In works such as \cite{freisleben1996genetic,osaba2018discrete,osaba2016improved}, the ATSP is employed as a benchmarking problem along with the TSP for developing and testing evolutionary computation methods \cite{precup2019nature,chakraborty2017swarm} such as a Genetic Local Search Algorithm, Water Cycle Algorithm and Bat Algorithm, respectively. Further examples can be found in \cite{boryczka2019harmony} and \cite{svensson2020constant} opting for a Harmony Search method and a Constant-factor Approximation solver. Theoretical studies have also been published, as those in \cite{cirasella2001asymmetric,oncan2009comparative,frieze1982worst}. Furthermore, diverse real-world oriented variants of the ATSP have also been formulated, aiming at adapting the problem to logistic-related environments \cite{arigliano2019time,gharehgozli2020high,osaba2018solving}.

Like every problem within TSP family, the ATSP implies a huge optimization challenge due to its NP-Hard nature. This is the main reason since a myriad of intelligent solvers have been designed for solving this problem. Additionally, ATSP recurrently serves a benchmarking purpose. Embracing this philosophy, the main objective of this paper is not to solve the ATSP up to its optimality, but to provide the first ever approach in the literature for being tackled with QC devices.

Lastly, it is important to highlight that the QUBO formulation adopted in this research for the ATSP is inspired by the canonical ATSP defined by Feld et al. in \cite{feld2019hybrid}. We refer interested readers to that paper for obtaining additional details on its QUBO expression.

\section{Hybrid Quantum Computing - Tabu Search Algorithm}\label{sec:QTA}

As mentioned in the introduction, one of the main objectives of this papers is to extend the outcomes obtained in our previous work published in \cite{osaba2020hybrid}, where we introduced our hybrid quantum-classical QTA. Accordingly, we do not give deep details of the QTA, encouraging interested researchers to read \cite{osaba2020hybrid} for delving deeper into more specific aspects. 

However, it is important to highlight the key principles that guided the design of our QTA: i) to reduce the non-profitable calls to the QC device and ii) to have a greater control over the accesses to QC resources. Following these pillars, the QTA indirectly reaches a reduction in the involved economical cost, associated with the usage of quantum computational resources. These economical expenses usually constitute obstacles for researchers to contribute to this field\footnote{https://www.forbes.com/sites/forbestechcouncil/2020/09/09/three-steps-toward-becoming-a-quantum-capable-organization/?sh=10bbe7c7bb6a}.

\begin{figure}[t]
	\centering
	\includegraphics[width=1.0\hsize]{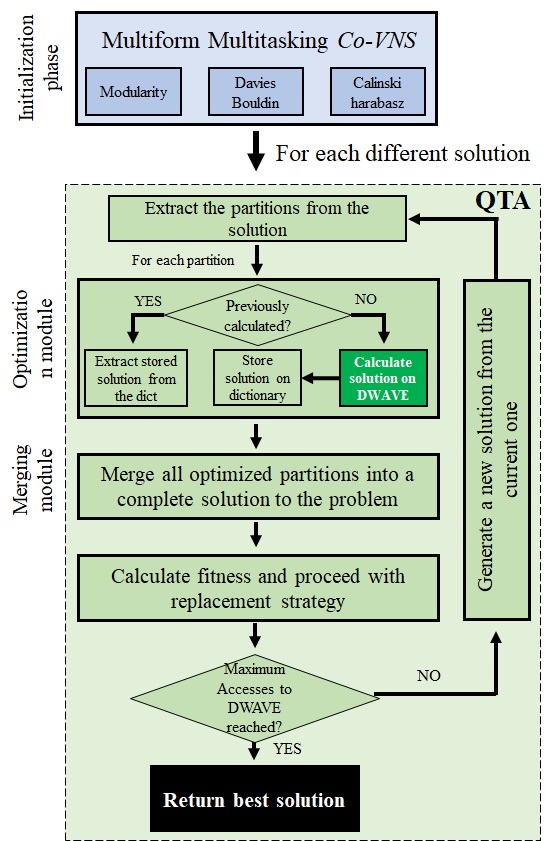}
	\caption{Workflow of QTA employed in this paper for the ATSP resolution.}
	\label{fig:QTA}
\end{figure}

Consequently, our QTA is divided into three different main steps for addressing large partitioning problems: i) the calculation of problem partitions, ii) the solving of the created subinstances through the use of D-Wave \textit{Advantage\_system1.1} and iii) the automatic merging of the independently-faced subproblems and the evaluation of the obtained unique global solution. We show the main architecture of QTA in Figure \ref{fig:QTA}. Furthermore, with the aim of properly understanding the main phases of QTA, we briefly describe the most important aspects of these stages.

\begin{itemize}
	\item \textit{Initialization procedure}: the initialization phase of the QTA is carried out by the above mentioned multiform CoVNS. In few words, CoVNS is a multitasking method composed by as much subpopulations as tasks to solve. Each of these subgroups is responsible for the solving of one sole task, accommodating the coevolutionary nature by means of periodical migrations of solutions between subpopulations. On this regard, and considering that the optimum graph partition does not compulsory entail a good decomposition scheme for the given optimization problem, we explore three different clustering solutions using three well-known metrics: Calinski Harabasz \cite{calinski1974dendrite}, Davies Boulding \cite{davies1979cluster} and Modularity \cite{newman2004finding}. Therefore, the multiform CoVNS offers the best partition found for each of these metrics. Lastly, if any practitioner decides not to employ any initialization problems, we suggest beginning the running of the technique with a randomly generated feasible solution.
	
	\item \textit{Optimization module}: first, note that the encoding employed for representing the clustering proposed for each ATSP dataset is the well-known label-based encoding \cite{hruschka2009survey}. Thus, the algorithm starts by extracting all partitions $\widetilde{\mathcal{G}}$ that compose the initial solution. Then, for each partition $\widetilde{\mathcal{G}^i}$, QTA calculates the route using the above mentioned D-Wave \textit{Advantage\_system1.1} device. Once the algorithm calculates the route of a specific subproblem, it stores this route into a \textit{Tabu Dictionary}, collecting in this way all previously considered clusters and their optimized solutions.
	
	In this way, QTA can resort to these previously calculated solutions in subsequent iterations of the search process, thus avoiding redundant accesses to the remote D-Wave machine. In other words, the \textit{Tabu Dictionary} is a strategy conceived to prevent the repeated and non-profitable calls to the QC cloud service for already calculated subgraphs, drastically reducing the amount of QC device accesses. We show in Figure \ref{fig:optimization} an illustration of this specific module using the ATSP as example problem.
	
	\begin{figure}[t]
		\centering
		\includegraphics[width=1.0\hsize]{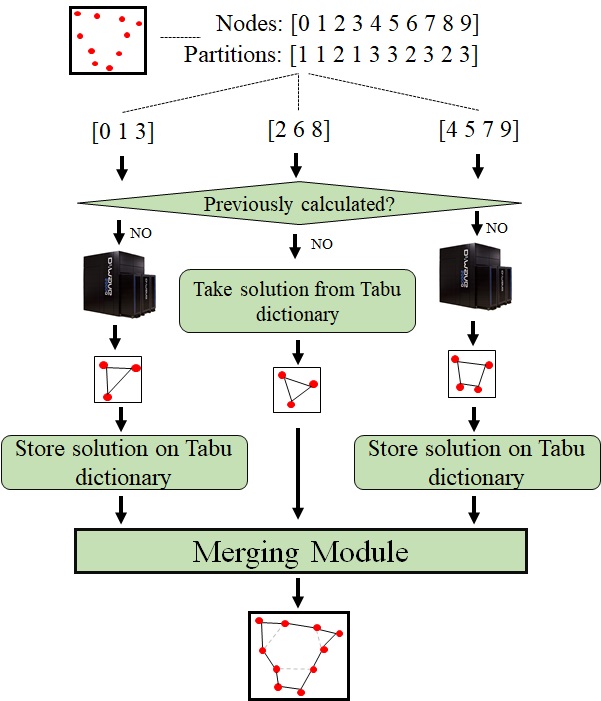}
		\caption{Optimization and Merging modules workflow using as example a 10-node ATSP dataset.}
		\label{fig:optimization}
	\end{figure}
	
	\item \textit{Merging module}: this module follows a greedy strategy seeking to compose the complete final solution $TSP^*$ taking as input the pool of $\widetilde{\mathcal{TSP}}$ subsolutions. In few words, $TSP^*$ is composed by all arcs that comprise each $TSP^{i}$ with the exception of those links needed to open up the closed loops-subroutes. The formation of the final cycle is conducted through the creation of $C$ bridges, each responsible for linking two different partitions in $\widetilde{\mathcal{G}}$. Once the merging is performed and the $TSP^*$ is built, QTA measures the total fitness using the cost based ATSP objective function. Then, the current solution is replaced if the new generated one represents an improvement. Finally, if the maximum amount of D-Wave calls has been reached, the QTA returns the best found solution. Otherwise, the current partition is evolved using an \textit{insertion function} \cite{osaba2016improved} and re-introduced in the optimization module.
\end{itemize}

\section{Experimental Setup and Results}\label{sec:exp}

As mentioned before, six different ATSP instances have been employed for this experimentation, each one belonging to the TSPLIB Benchmark and composed by 17 to 48 nodes.  More concretely, chosen instances are \texttt{br17}, \texttt{ftv33}, \texttt{ftv35}, \texttt{ftv38}, \texttt{p43} and \texttt{ry48p}. Each instance have been executed 20 times in order to obtain statistically representative results. Furthermore, and due to the stringent limits on the access to the QC device, we have also made use of the local QBSolv alternative offered by D-Wave. This way, we have been able to conduct a representative experimentation comprising instances up to 43 nodes. Thus, and embracing the same strategy as remarkable papers such as \cite{feld2019hybrid,teplukhin2020electronic}, the default setup of QBSolv has been kept\footnote{At the time of writing, QBSolv's current version is 0.3.2.}.

Additionally, the maximum clusters size for the QTA has been established on 10. Because of the QUBO matrix-based representation requirements and the 5436 qubits available, this number is the maximum size that the current \textit{D-Wave Advantage\_system1.1} can assume. Furthermore, 40 calls to QC hardware have been defined as termination criterion. 

With the aim of contributing to the TSP and ATSP community, as well as to QC research field, we openly offer the QTA code described in this paper. Thus, the Python implementation of our method, together with the scripts needed for solving ATSP instances, is publicly available in github\footnote{https://github.com/EnekoOsaba/QAT4ATSP}. It is also worth mentioning that the main algorithmic base used for developing our method is also available in GitHub, openly offered by user \textit{mstechly}\footnote{https://github.com/BOHRTECHNOLOGY/quantum\_tsp}.

Last but not least, it is important to bring to the fore that the main objective of our QTA is to obtain promising outcomes while overcoming one of the major concerns voiced by the QC community. Thus, our motivations with this research are a) to offer practitioners and researchers a fully configurable method which provides absolute control to QC remote accesses, and b) to drastically reduce the number or non-profitable calls to the quantum resources, avoiding possible connection problems and undesirable waste of the committed quantum resource budget.

With this in mind, Table \ref{tab:Results} shows the results (average/standard deviation/best) yielded by both QTA and QBSolv. We have also added to this table the difference between the best performing method and the runner-up represented by a percentage. Additionally, for faithfully measuring the key objective of our QTA in terms of the reduction of QC device accesses, Table \ref{tab:Results} also contemplates the average calls needed by the QBSolv technique for achieving each outcome.

\begin{table}[h!]
	\centering
	\caption{Obtained results (average/standard deviation/best) using QTA and QBSolv. Best results have been highlighted in bold.}
	\label{tab:Results}
	\resizebox{1.0\columnwidth}{!}{
		\begin{tabular}{| c | r r r | r r r | r|}
			\hline   
			\multicolumn{1}{|c|}{} & \multicolumn{3}{c|}{QTA} & \multicolumn{4}{c|}{QBSolv}\\
			\hline
			Instance & Avg & Std & Best & Avg & Std & Best & AvAcc\\ 
			\hline
			\texttt{br17} & \textbf{39.0} & 0.0 & 39 & \textbf{39.0} & 0.0 & 39 & 184.0 \\
			\texttt{ftv33} & \textbf{1576.3} & 51.94 & 1487 & 1584.4 & 43.16 & 1514 & 564.2\\
			\texttt{ftv35} & \textbf{1782.4} & 59.72 & 1686 & 1819.5 & 44.67 & 1749 & 585.5\\
			\texttt{ftv38} & \textbf{1848.4} & 74.41 & 1719 & 1908.2 & 50.34 & 1790 & 784.8\\
			\texttt{p43} & 5785.4 & 63.82 & 5698 & \textbf{5703.4} & 10.04 & 5686 & 928.8\\
			\texttt{ry48p} & 19456.5 & 503.0 & 18302 & \textbf{17873.0} & 374.9 & 17401 & 1036.6\\
			\hline
		\end{tabular}
	}
\end{table}

Several conclusion can be drawn by analyzing the results depicted in Table \ref{tab:Results}. First, it can be confirmed that both algorithms perform similarly. On the one hand, QTA reaches better results in 3 out of 6 datasets (\texttt{ftv33}, \texttt{ftv35} and \texttt{ftv38}), whereas QBSolv slightly outperforms QTA in two of the considered instances (\texttt{p43} and \texttt{ry48p}). Finally, both solving schemes work equal in \texttt{b17} instance. In any case, performance gaps are certainly non-significant. This situation brings, indeed, a great advantage for our introduced QTA. This is so since it can obtain similar results using remarkably less QC resources. Specifically, QTA just requires 40 calls to D-Wave's QC device, while QBSolv shows a considerably increasing demand of QC accesses, ranging from 184.0 calls for the smallest instance to 1036.6 for the bigger one. This aspect, together with the competitiveness in terms of results, evinces that our method provides a considerable advantage. These results help us to confirm that QTA is a novel classical-quantum hybrid solver scheme which offers a good performance for dealing with partitioning problems ensuring a less demanding behavior in terms of QC resource calls.

\section{Conclusions and Future Work}\label{sec:conc}

This paper has been devoted to extending the overall findings drawn in our previously published work \cite{osaba2020hybrid}, focused on the design and implementation of a quantum-classical hybrid optimization method coined as Hybrid Quantum Computing - Tabu Search Algorithm (QTA). To do that, we have adapted our method to the well-known Asymmetric Traveling Salesman Problem, representing the first efforts for tackling this problem through the quantum computing paradigm. Thus, for measuring the quality of our method on the ATSP, we have employed 6 different well-known TSPLIB instances, made up by 17 to 48 nodes. Experimental outcomes confirm the results of our previous work, illustrating that our developed QTA is a promising method for solving partitioning problems. As future work, we have planned the application of the QTA for larger problems, closer to real-world contexts. Extensive research on the merging module of QTA will be also undertaken by covering the application of quantum-based solvers in this stage.

\section*{Acknowledgments}

Eneko Osaba, Esther Villar-Rodriguez and Izaskun Oregui would like to thank the Basque Government for its funding support through the EMAITEK and ELKARTEK (Quantek Project) programs.

\bibliographystyle{ACM-Reference-Format}
\bibliography{gecco} 

\end{document}